\def\spose#1{\hbox to 0pt{#1\hss}}

\def\multleft#1{\hbox to size{\vbox {\halign {\lft{##}\cr #1}}\hfill}\par}
\def\multright#1{\hbox to size{\vbox {\halign {\rt{##}\cr #1}}\hfill}\par}

\def\today{\ifcase\month\or January\or February\or March\or April\or May\or
      June\or July\or August\or September\or October\or November\or December\fi
      \space\number\day, \number\year}
\def\Msun{\hbox{$\rm\thinspace M_{\odot}$}}

\def\H2{\hbox{H$_{2}$}}

\newcommand{\gtsim}{\mbox{{\raisebox{-0.4ex}{$\stackrel{>}{{\scriptstyle\sim}}
$}}}}

\documentclass{mn2e}

\usepackage{times}
\usepackage{amssymb}
\usepackage{epsfig}

\begin{document}
\hsize=6truein
       
\title[The quasar radio luminosity - black hole mass relation]
{The relationship between radio luminosity and black-hole mass in
optically-selected quasars}

\author[R.J.~McLure and M.J~Jarvis]
{Ross J. McLure$^{1}$\thanks{Email: rjm@roe.ac.uk} and Matt
J. Jarvis$^{2}$\thanks{Email: mjj@astro.ox.ac.uk}\\
\footnotesize\\
$^{1}$Institute for Astronomy, University of Edinburgh, 
Royal Observatory, Edinburgh, EH9 3HJ, UK\\
$^{2}$Astrophysics, Department of Physics, Keble Road, Oxford OX1 3RH}

\maketitle
     
\begin{abstract}
Using a sample of more than 6000 quasars from the Sloan digital sky
survey (SDSS) we compare the black-hole mass distributions of
radio-loud and radio-quiet quasars. Based on the 
virial black-hole mass estimator the radio-loud quasars (RLQs) are
found to harbour systematically more massive black holes 
than radio-quiet quasars (RQQs) with very 
high significance ($\gg 99.99\%$), with mean black-hole masses of 
$<\log(M_{bh}/\Msun)>=8.89\pm0.02$ and 
$<\log(M_{bh}/\Msun)>=8.69\pm0.01$ for the RLQs and RQQs respectively.
Crucially, the new RLQ and RQQ samples have indistinguishable 
distributions on the redshift-optical luminosity plane, excluding 
the possibility that either parameter is responsible for the observed 
black-hole mass difference. Moreover, this black-hole mass 
difference is shown to be in good
agreement with the optical luminosity difference observed between RLQ
and RQQ host galaxies at low redshift 
(i.e. $\Delta M_{host}=0.4-0.5$ magnitudes). Within the 
SDSS samples, black-hole mass is strongly correlated with both radio
luminosity and the radio-loudness $\mathcal{R}$ parameter 
($>7\sigma$ significance), although the range in radio luminosity at a
given black-hole mass is several orders of magnitude. It is therefore
clear that the influence of additional physical parameters or
evolution must also be invoked to explain the quasar radio-loudness dichotomy.
\end{abstract}

\begin{keywords}
black hole physics - galaxies:active - galaxies:nuclei - quasars:general  
\end{keywords}

\section{INTRODUCTION}
Ever since the discovery that only a small minority of
optically-selected quasars are also luminous radio sources 
(e.g. $L_{\rm{5GHz}}>10^{24}$WHz$^{-1}$sr$^{-1}$), many studies have attempted 
to isolate the physical mechanism underlying this so-called 
quasar radio-loudness dichotomy. Although previous studies found a clear 
bimodality in the radio luminosities of optically-selected 
quasars (eg. Kellermann et al. 1989; Miller, Peacock \& Mead 1990), more
recently the very existence of the radio-loudness dichotomy has been questioned
(Lacy et al. 2001; Cirasuolo et al. 2003; although see 
Ivezi\'{c} et al. 2002 for
an alternative viewpoint). The principal reason behind this renewed
interest in the radio properties of optically selected quasars is
the ability to combine the large SDSS and 2dF optical quasar samples with
wide-area radio surveys such as the Faint Images
of the Radio Sky at Twenty-cm (FIRST) and the NRAO VLA Sky
Survey (NVSS). The FIRST survey (Becker, White \& Helfand 1995) 
in particular has identified large numbers of
so-called radio-intermediate quasars which have largely filled-in the 
apparent gap in radio luminosities between the RLQ and RQQ populations
(eg. Lacy et al. 2001). Consequently, the distribution of optically-selected quasars on the optical-radio luminosity plane is undoubtedly
more continuous than was previously thought. Irrespective of
this, the fundamental question of what causes
luminous quasars with seemingly identical optical properties to 
differ in their radio luminosities by several orders of magnitude
remains unanswered.

Recent progress has been made in largely 
eliminating two parameters which were originally suspected of
influencing the radio-loudness dichotomy; host-galaxy morphology and 
cluster environment. Thanks largely to the 
Hubble Space Telescope (HST), quasar host-galaxy morphologies have 
now been investigated out to intermediate redshifts, $0.1<z<0.5$ 
(eg. Dunlop et al. 2003; Schade et al 2000; McLure et al. 1999; 
Disney et al. 1996).
Drawing together the results of these studies, it is clear that 
the hosts of optically luminous quasars (i.e. $M_{R}<-24$) are 
bulge-dominated, spheroidal galaxies irrespective of radio luminosity 
(Dunlop et al. 2003;  Schade et al. 2000). In the light of the 
discovery of the correlation between black-hole and bulge mass
(Magorrian et al. 1998; Gebhardt et al. 2000; Ferrarese \& Merritt
2000), this result is perhaps not surprising. However, it should 
also be remembered that in the optical the hosts of RLQs are 
consistently found to be $\simeq 0.5$ magnitudes brighter than their 
RQQ counterparts (eg. Dunlop et al. 2003), the implications of which
will be discussed in Section 3. In addition, it now appears that 
cluster environment does not greatly 
influence quasar radio luminosities either. Recent studies of the cluster
environments of AGN have tended to show that quasars prefer to 
inhabit poor clusters (e.g. Abell class 0) or groups, rather than 
the cores of rich clusters (Best 2004; S\"{o}chting, Clowes \&
Campusano 2004). Where a direct 
comparison of matched samples of RLQs and RQQs has been performed 
(eg. Wold et al. 2001; McLure \& Dunlop 2001a) little difference has
been found in their respective cluster environments.

Following the inconclusive nature of the results regarding the Mpc and
Kpc-scale environments of RLQs and RQQs, attention has now become
focussed on possible differences in the properties of their central
engines. At the same time the robustness of the so-called
virial black-hole mass estimator (Wandel, Peterson \& Malkan 1999; 
Kaspi et al. 2000) has now become established, allowing estimates of
AGN black-hole masses to be derived from a continuum
luminosity and broad-line FWHM measurement only (i.e. $M_{bh}\propto \lambda
L_{\lambda}^{0.6}$FWHM$^{2}$). Consequently, many studies have exploited the
virial mass estimator to investigate whether black-hole mass could be 
the hidden parameter controlling the radio-loudness dichotomy; with
widely contrasting results. Several authors
(eg. Laor 2000; McLure \& Dunlop 2001b,2002; Dunlop et al. 2003) have
concluded that RLQs have systematically higher black-hole masses than
their RQQ counterparts, and have suggested the existence of a black-hole-mass threshold (e.g. $M_{bh}\simeq 10^{8}\Msun$) above which the
fraction of RLQs increases significantly. This separation of the RLQ and
RQQ populations in terms of black-hole mass is further supported by the 
principal components analysis of RLQ and RQQ spectra by Boroson (2002).
Moreover, by combining the FIRST Bright Quasar Survey (FBQS, Gregg
et al. 1996) and the PG quasar survey (Green, Schmidt \& Liebert
1986), Lacy et al. (2001) found that, given enough dynamic range, 
radio luminosity and black-hole mass were strongly correlated, 
albeit with substantial scatter. More recently, McLure et al. (2004) 
found a significant correlation between 
black-hole mass and integrated low-frequency (151 MHz) radio luminosity
from a complete sample of 41 powerful radio galaxies at $z\simeq0.5$.

In contrast to the previously mentioned studies, several authors have
concluded that no strong link exists between black-hole mass and radio
luminosity in AGN (eg. Urry 2003; Oshlack et al. 2002; Woo \& Urry
2002; Ho 2002). With particular relevance to optically luminous
quasars, both Oshlack et al. (2002) and Woo \& Urry (2002) highlight 
that the apparent existence of extremely radio-loud flat-spectrum quasars
(FSQs), with virial black-hole mass estimates in the 
range $10^{6}\Msun<M_{bh}<10^{8}\Msun$, effectively rule out a strong
correlation between black-hole mass and radio-loudness (although 
see Jarvis \& McLure 2002 for a different interpretation of the FSQ data). 

In this paper we revisit the issue of whether black-hole mass and
radio luminosity are related in quasars using the virial black-hole mass
estimates calculated by McLure \& Dunlop (2004) for objects from the 
SDSS quasar catalog II (Schneider et al. 2003). The aim of this paper
is to determine whether RLQs and RQQs, matched in terms of redshift and 
optical luminosity, have distinguishable black-hole mass distributions
or not. Throughout the paper we adopt the spectral index 
convention $f_{\nu} \propto
\nu^{-\alpha}$ and the following cosmology: 
$H_{0}=70$km s$^{-1}$, $\Omega_{\Lambda}=0.7$, $\Omega_{m}=0.3$.

\section{The Sample}
\begin{figure*}
\centerline{\epsfig{file=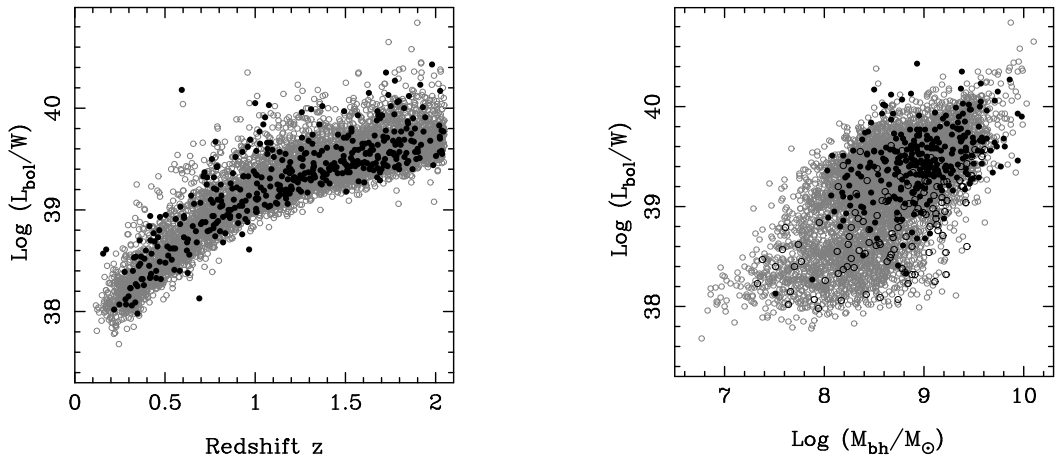,width=14.0cm,angle=00}}
\caption{(a) Bolometric luminosity versus redshift for the
RQQ (open grey circles) and RLQ (filled black circles) samples.(b) 
Bolometric luminosity versus virial black-hole mass estimate for the
same samples, with those RLQs with
$L_{5\rm{GHz}}<10^{24}$WHz$^{-1}$sr$^{-1}$ plotted as open black circles.}
\label{fig1}
\end{figure*}

The sample adopted in this paper is drawn from the SDSS quasar catalog
2 (Schneider et al. 2003) which consists of 16714 quasars in the
redshift interval $0.1<z<5.3$ with $M_{i}(AB)<-22$. The black-hole
mass estimates are taken from McLure \&
Dunlop (2004) who produced virial black-hole mass estimates for more
than 12000 of the SDSS quasar catalog 2 (SQC2) objects with redshifts $z~\le
~2.1$, based on both the H$\beta$ and MgII
emission lines. Full details of the modelling of the SDSS quasar
spectra and the calibrations of the virial mass estimator can
be found in McLure \& Dunlop (2004) and McLure \& Jarvis (2002). The 
final samples of RLQs and RQQs were selected from the full SQC2 
via application of the following criteria:
\begin{enumerate}
\item{All objects which fell outside the SDSS/FIRST overlap region
were excluded to ensure consistent radio data for the final sample.}

\item{All objects with $i(AB)>19.1$ or which did not satisfy the SDSS 
quasar colour selection algorithm for follow-up spectroscopy were also 
excluded. This criterion ensures that the final samples are not biased 
by the inclusion of pre-selected FIRST sources 
or serendipitous targets (Schneider et al. 2003).}
\end{enumerate}

The remaining objects were cross-correlated
with the McLure \& Dunlop (2004) black-hole mass estimates which cover
$>90\%$ of the SDQ2 objects with $z<2.1$. This final list of quasars
was then separated into the 
RLQ and RQQ samples based on the radio-loudness $\mathcal{R}$ 
parameter ($\mathcal{R}=f_{5GHz}/f_{B}$, where $f_{5GHz}$
\footnote{The 5GHz flux-density is extrapolated from 1.4GHz using 
the mean spectral index of
FBQS quasars; $\alpha=0.5$ (White et al. 2000)} and $f_{B}$
are the  observed flux densities at 5GHz and 4400\AA\ respectively). 
Following the standard convention we 
classify every object with $\mathcal{R}\ge10$ as radio-loud, and everything 
with $\mathcal{R}<10$ as radio-quiet. An  upper limit in $\mathcal{R}$ was
calculated for quasars undetected by FIRST using the nominal FIRST 
object-detection threshold of 1mJy. To investigate the possibility
that the quasar radio luminosities are underestimated, due to the FIRST
survey resolving out extended flux (e.g. Blundell 2003), we
cross-correlated our samples with the lower resolution NVSS
survey. The results of this showed no significant differences in
1.4GHz flux densities, which suggests that the radio emission of 
these optically-selected quasars is fairly compact.

The final RQQ and RLQ samples
produced by our selection process comprise 6099 and 436 objects respectively. 
We note here that adopting $\mathcal{R}\ge10$ as our 
radio-loudness threshold is conservative, in the sense that $30\%$ of
the final RLQ sample have 5GHz radio luminosities less than the alternative
radio-luminosity based threshold of $10^{24}$WHz$^{-1}$sr$^{-1}$
(Miller, Peacock \& Mead 1990). The results of adopting the strict
radio luminosity threshold are discussed further in Section 3.

\subsection{Matched samples}

In order to investigate any differences in the 
black-hole mass distributions of the RLQs and RQQs the samples should be
free from biases which could affect the virial
black-hole mass estimator. Given the role of optical luminosity in the
virial mass estimator ($M_{bh}\propto \lambda
L_{\lambda}^{0.6}$FWHM$^{2}$), this essentially requires that the two samples
are indistinguishable in terms of their optical luminosity and 
redshift distributions. 

In Fig \ref{fig1}a we show the RLQ and 
RQQ samples on the $L_{bol}-z$ plane where, following McLure \& Dunlop
(2004), the quasar bolometric luminosities are estimated from either the
5100\AA\, or 3000\AA\, luminosities as $L_{bol}=9.8\lambda L_{5100}$
or $L_{bol}=5.9\lambda L_{3000}$, depending on
their redshift (McLure \& Dunlop 2004 demonstrated that the ratio of
these two bolometric luminosity estimators is unity for $\simeq1200$
quasars where both could be measured). The application of the one-dimensional 
Kolmogorov-Smirnov (KS) test finds that both the $L_{bol}$ and redshift
distributions of the RLQ and RQQ samples are indistinguishable;
$p=0.16$ and $p=0.50$ respectively. In addition, an application of the 
two-dimensional KS test to the $L_{bol}-z$ distribution shown in Fig
\ref{fig1}a confirms that the two-dimensional distributions of the RLQ and
RQQ samples are also indistinguishable; $p=0.21$. Consequently, any 
black-hole mass difference between the two quasar 
samples should not be a product of significant biases in their redshift or 
optical luminosity distributions.

\section{Black-hole mass distributions}

The distributions of the RQQ and RLQ samples on the $L_{bol}-M_{bh}$
plane are shown in Fig \ref{fig1}b. The application of the
two-dimensional KS test returns a probability of only $p=1.2\times
10^{-10}$ that both samples are drawn from the same parent population.
The mean black-hole masses for the RQQ and RLQ samples are as follows:
\begin{displaymath}
{\rm RLQ:}\,\,\, <\log(M_{bh}/\Msun)>=8.89\pm0.02 \,\,\,\,\,(8.93)
\end{displaymath}
\begin{displaymath}
{\rm RQQ:}\,\,<\log(M_{bh}/\Msun)>=8.69\pm0.01 \,\,\,\,\,(8.74)
\end{displaymath}
\noindent
where the quoted uncertainties are the standard errors and
the values in parenthesis are the sample medians. Although the
optical luminosity distributions of the RLQ and RQQ samples are
indistinguishable, the mean optical luminosity of the RLQs is
0.06 dex brighter than that of the RQQs. Given that the virial
mass estimator adopted here has a luminosity dependence of
$\lambda L_{\lambda}^{0.6}$ (McLure \& Dunlop 2004), this will account
for 0.04 dex of the black-hole mass difference between the samples. 
Consequently, accounting for this slight luminosity bias, we conclude 
that RLQs typically harbour black holes 0.16 dex (45\%) more massive than 
their radio-quiet counterparts. Moreover, by calculating the mean
RLQ$-$RQQ black-hole mass difference in narrow luminosity bins (0.3 dex
width) it was confirmed that the black-hole mass difference is consistent
with a constant 0.16 dex off-set over the full luminosity range.

\subsection{Line-width difference}
The measured difference in mean black-hole mass between the RLQ and
RQQ samples is predominantly due to fact that the FWHM of the low-ionization
H$\beta$ and MgII emission lines of the RLQs are $\simeq 20\%$ larger than
in the RQQs of the same optical luminosity. The conclusion that RLQs
exhibit broader low-ionization emission lines than RQQs is in good
agreement with many previous studies (eg. Boroson \& Green 1992;
Corbin 1997; McLure \& Dunlop 2002; Sulentic et al. 2003). However, although
the velocity-width difference between RLQs and RQQs has been noted 
previously, because of potential selection effects, attributing 
this directly to a difference
in black-hole mass has been problematic. For example, given that the
FWHM of low-ionization emission lines in RLQs are known to be influenced by
viewing angle (Wills \& Browne 1986; Brotherton et al. 1996),
the velocity-width difference between RLQs and RQQs could potentially
be an orientation effect produced by comparing radio and optically
selected samples. However, as discussed in Section 2, here both samples 
have been selected consistently by the SDSS $ugri$ colour algorithm,
with targeted FIRST sources deliberately excluded. 
Combined with the excellent sample matching in terms of optical
luminosity and redshift, the influence of potential orientation
effects should therefore be minimized. 
Consequently, the simplest interpretation of these new
results is that at a fixed optical luminosity RLQs do, on average,
harbour more massive black-holes than RQQs, accreting at a 
commensurately lower fraction of their Eddington limit.

\subsection{Comparison with host-galaxy results}

\begin{figure*}
\centerline{\epsfig{file=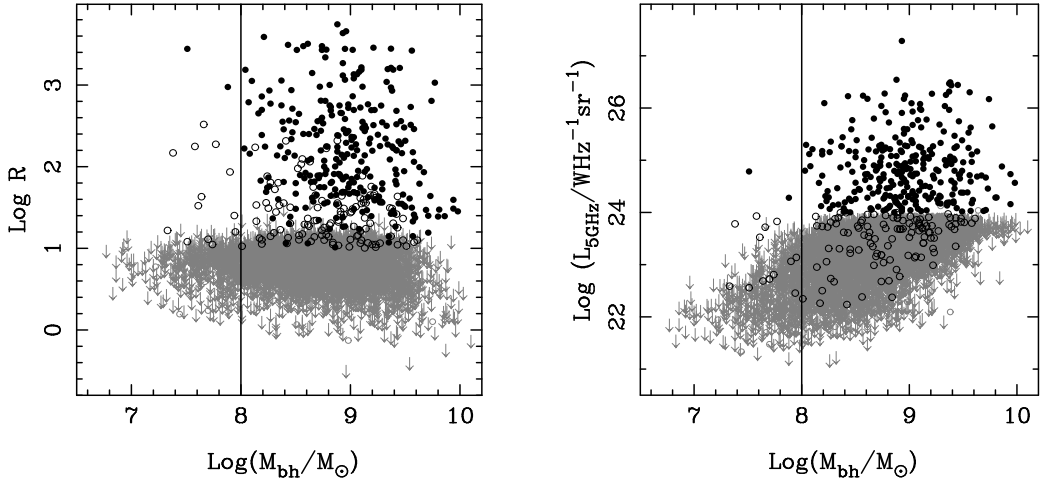,width=13.8cm,angle=0}}
\caption{(a) Radio-loudness parameter $\mathcal{R}$ 
versus black-hole mass. (b) Absolute 5GHz radio luminosity versus 
black-hole mass. Symbols as Fig 1. RQQs undetected by FIRST are shown as 
upper limits. A black-hole mass of $10^{8}\Msun$ is 
highlighted by a solid vertical line in both figures.}
\label{fig2}
\end{figure*}
Irrespective of the virial black-hole mass estimates presented here,
it could reasonably be argued that the correlation between 
bulge and black-hole mass ($M_{bh}-M_{bulge}$) and the well known $K-z$ 
relation for powerful radio galaxies (Lilly \& Longair 1984)
immediately imply that RLQs should harbour black holes with masses
$\gtsim 10^{8}\Msun$. The recent Willott et al. (2003) 
study showed that the radio-galaxy $K-z$ relation is consistent 
with the passive evolution of a $\simeq 3L^{\star}$ elliptical formed
at high redshift
($z\geq5$). Therefore, if the predictions of
orientation-based unification are correct, the host-galaxies of 
RLQs of comparable radio luminosity should have similar $K-$band 
luminosities . 

Notably, using the McLure \& Dunlop (2002) fit to the $M_{bh}-M_{bulge}$
relation, the mean black-hole mass of the RLQ sample corresponds to
a $K-$band bulge luminosity of $3.8\pm0.5 L^{\star}$
(assuming a $z=0$ colour of $R-K=2.7$; Bruzual \& Charlot 2003), in
good agreement with the Willott et al. $K-z$ relation. Moreover, it is
also interesting to compare the RLQ black-hole mass
estimates derived here with those derived via host-galaxy luminosities for a
complete sample of 41 $z\simeq 0.5$ radio galaxies by 
McLure et al. (2004). Based on HST imaging of radio galaxies 
spanning the same range of radio luminosities 
as the current RLQ sample, the mean black-hole mass was determined to be
$<\log (M_{bh}/\Msun)>=8.87\pm0.04$, in excellent agreement with the
mean black-hole mass of the RLQ sample derived here.

In addition to comparisons with radio galaxy studies, it is also
possible to compare the black-hole mass difference between the RLQ and 
RQQ samples determined here with imaging studies of 
quasar host-galaxies.
The most accurate quasar host-galaxy studies are those based on 
high-resolution HST imaging data (eg. Dunlop et
al. 2003; Schade et al. 2000; Disney et al. 1995). Among the
various HST-based host galaxy studies, the Dunlop et al. (2003) study 
has the advantage of featuring RLQ and RQQ samples which
have well matched redshift - optical luminosity distributions. 
Based on the results of this $R-$band imaging study, the hosts of RLQs 
are typically $\simeq 0.4$ mags
brighter than their RQQ counterparts. Combining this with the McLure
\& Dunlop (2002) fit to the $M_{bh}-M_{bulge}$ relation implies
that RLQ and RQQ black-hole masses should differ by $\simeq 0.2$ dex,
in good agreement with the virial-based results derived here.

\section{The correlation between black-hole mass and radio loudness}

In Fig \ref{fig2} we plot both the radio-loudness $\mathcal{R}$ 
parameter and absolute radio luminosity versus black-hole mass for 
the RLQ and RQQ samples. In order to assess the significance of the apparent
correlations in the presence of upper limits we have applied 
the generalized Kendall's $\tau$ test in the ASURV package (Isobe,
Feigelson \& Nelson 1986) . The Kendall's $\tau$ test detects a 
highly significant correlation in both cases, returning values of
$\tau=0.0497\,\,(7.6\sigma)$ and $\tau=0.0697\,\, (11.4\sigma)$ for 
the $\mathcal{R}-M_{bh}$ and $L_{5GHz}-M_{bh}$ correlations
respectively. 

The detection of a significant
correlation between black-hole mass and radio-loudness is in good
agreement with the Lacy et al. (2001) study of the combined FBQS+PG sample. 
Furthermore, it can be seen from Fig \ref{fig2}a that the RLQs are 
virtually exclusively confined to $M_{bh}\geq 10^{8}\Msun$, as
previously found by Laor (2000), Lacy et al. (2001), Boroson (2002)
and McLure \& Dunlop (2002). Qualitatively, the fraction of RQQs with
$M_{bh}\leq 10^{8}\Msun$ is $9.2\% \pm 0.4$, whereas the
equivalent fraction of RLQs is only $3.7\% \pm 1.0$. 
In fact, if the definition of radio-loud is restricted to only those
quasars with $L_{5GHz}>10^{24}$WHz$^{-1}$sr$^{-1}$ only two objects have 
$M_{bh}\leq 10^{8}\Msun$ ($<1\%$). This result, combined with the
radio galaxy $K-z$ relation (e.g. Jarvis et al. 2001; Willott et al. 2003), 
strengthens the claim that regardless of the exact nature of the 
black-hole mass - radio luminosity relationship in 
luminous quasars, sample selection based on high radio luminosity is 
effective in isolating the largest black holes, and presumably 
host galaxies, at all epochs (eg. McLure 2003). 

\subsection{Flat-spectrum quasars} 

In contrast to the results found here, we note that Woo \& Urry (2002)
found no indication of a  $\mathcal{R}-M_{bh}$
correlation in their study of a heterogeneous sample of 747 
quasars in the redshift interval $0<z<2.5$. The simple reason for this
apparent discrepancy is that within their sample Woo \& Urry include
the 39 flat-spectrum quasars (FSQs) studied by Oshlack et al. (2002).
Jarvis \& McLure (2002) have examined the likely effects of 
orientation on both the radio luminosities and black-hole mass
estimates for the Oshlack et al. (2002) sample. Although estimates
of the intrinsic unbeamed $\mathcal{R}$ parameters are hard to
accurately determine, Jarvis \& McLure concluded that the radio
luminosities and black-hole mass estimates of Oshlack et al. (2002)
are likely to be over/under-estimated by factors of $\simeq 100$ and
$\simeq 4$ respectively. 

Consequently, in all likelihood the 
Oshlack et al. FSQs are not inconsistent with 
the $L_{\rm{5GHz}}-M_{bh}$ correlation seen in
Fig \ref{fig2}b, with all the FSQs with {\it intrinsic} radio luminosities
$L_{\rm{5GHz}}>10^{24}$WHz$^{-1}$sr$^{-1}$ likely to have $M_{bh}\,\gtsim\,
10^{8}\Msun$. Notably, if the FSQs are removed from the 
Woo \& Urry (2002) sample, then their results 
are entirely consistent with those derived here.

\section{Conclusions}

In this paper we have compared the black-hole mass distributions of 
RQQs and RLQs using large samples which have indistinguishable 
distributions in the redshift-optical luminosity plane. The results of
this comparison demonstrate that at a fixed optical luminosity RLQs
harbour black-hole masses which are typically 0.16 dex (45\%) more
massive than their radio-quiet counterparts. In agreement with
previous studies, this black-hole mass difference is shown to be
largely produced by RLQs displaying broader low-ionization emission 
lines than comparable RQQs. Moreover, using the combined RQQ+RLQ
sample we have shown that black-hole mass is strongly correlated with
both absolute radio luminosity and the radio-loudness parameter
$\mathcal{R}$. Furthermore, the black-hole masses of genuine
RLQs are found to be virtually exclusively confined to $M_{bh}\geq
10^{8}\Msun$, in good agreement with the conclusions of previous studies
(Laor 2000; Boroson 2002, McLure \& Dunlop 2002). 

Combining the virial black-hole mass estimates with the 
local $M_{bh}-M_{bulge}$ relation suggests that RLQ host galaxies are
fully consistent with the radio galaxy $K-z$ relation, in agreement 
with AGN orientation-based
unification. It is therefore 
clear that sample selection based on high radio luminosity is 
an effective method for cleanly isolating the largest black-holes, and 
presumably host galaxies, at all epochs. However, the new results 
presented here also confirm that no tight 
correlation exists between radio luminosity and black-hole mass, and
that at a fixed black-hole mass the range in quasar radio luminosities covers 
several orders of magnitude. Consequently, there is clearly a large
overlap between the RLQ and RQQ populations, with many RQQs being
indistinguishable from RLQs in terms of black-hole mass, accretion
rate, orientation, host galaxy properties and cluster environment. In
conclusion, it therefore appears inescapable that the influence of
additional physical parameters such as black-hole spin and/or
evolutionary effects must contribute to the radio-loudness dichotomy.

\section{acknowledgments}
RJM and MJJ are funded by PPARC PDRAs. This publication makes use of 
the material provided in the 
FIRST and SDSS DR1 surveys. FIRST is funded by the
National Astronomy Observatory (NRAO) and is a research facility of
the U.S. National Science foundation and uses the NRAO Very Large
Array. Funding for the creation and distribution of the SDSS Archive
has been provided by the Alfred P. Sloan Foundation, the Participating
Institutions, the National Aeronautics and Space Administration, the
National Science Foundation, the U.S. Department of Energy, the
Japanese Monbukagakusho, and the Max Planck Society. Further details
of the SDSS survey can be found on http://www.sdss.org/.

\end{document}